\documentclass[a4paper,11pt,twoside]{article}

\pdfoutput=1

\makeatletter

\usepackage{amsmath,amssymb}

\usepackage[english]{babel}

\usepackage{cancel}
\usepackage{color}

\usepackage{graphicx}
\usepackage{geometry}

\usepackage{hyperref}

\usepackage{url}

\numberwithin{equation}{section}
\allowdisplaybreaks[1]
\newcommand{\be}{\begin{equation}}
\newcommand{\ee}{\end{equation}}

\newcommand\duf{\star_{\mathcal{D}}}
\newcommand\bbone{{ \mathbb{I}}}

\DeclareMathOperator{\tr}{Tr}

\newcommand{\labitem}[2]{\def\@itemlabel{\textbf{#1}}\item\def\@currentlabel{#1}\label{#2}}

\newcommand{\institute}[1]{\newcommand{\@institute}{#1}}

\renewcommand{\maketitle}{
\vspace*{0.5\baselineskip}
{
\center\LARGE\noindent\@title\par
}%
\vspace{1.5\baselineskip}
{
\center\normalsize\noindent\ignorespaces\@author\par
}%
\vspace{0.5\baselineskip}
{
\center\normalsize\ignorespaces\@institute\par
}%
\vspace{2\baselineskip}
}%

\renewenvironment{thebibliography}[1]{%
\section*{References}%
\frenchspacing\small%
\begin{list}{[\arabic{enumi}]}%
{%
\usecounter{enumi}\parsep=2pt\topsep 0pt%
\settowidth{\labelwidth}{[#1]}%
\leftmargin=\labelwidth\advance\leftmargin\labelsep%
\rightmargin=0pt\itemsep=1pt\sloppy%
}%
}{\end{list}}

\begin{document}

\title{Closed star product on noncommutative $\mathbb{R}^3$ \\
and scalar field dynamics}

\author{Tajron Juri\'c$^a$, Timoth\'e Poulain$^b$ and Jean-Christophe Wallet$^b$}

\institute{%

\textit{$^a$Ru\dj er Bo\v{s}kovi\'c Institute, Theoretical Physics Division\\
Bijeni\v{c}ka c.54, HR-10002 Zagreb, Croatia}\\
e-mail:\href{mailto:tjuric@irb.hr}{\texttt{tjuric@irb.hr}}\\[1ex]%

\textit{$^b$Laboratoire de Physique Th\'eorique (UMR8627)\\
CNRS, University of Paris-Sud, University of Paris-Saclay, 91405 Orsay, France}\\
e-mail:\href{mailto:timothe.poulain@th.u-psud.fr}{\texttt{timothe.poulain@th.u-psud.fr}}, \href{mailto:jean-christophe.wallet@th.u-psud.fr}{\texttt{jean-christophe.wallet@th.u-psud.fr}}\\[1ex]%
}%

\date{\today}

\maketitle


\begin{abstract} 
We consider the noncommutative space $\mathbb{R}^3_\theta$, a deformation of $\mathbb{R}^3$ for which the star product is closed for the trace functional. We study one-loop IR and UV properties of the 2-point function for real and complex noncommutative scalar field theories with quartic interactions and Laplacian on $\mathbb{R}^3$ as kinetic operator. We find that the 2-point functions for these noncommutative scalar field theories have no IR singularities in the external momenta, indicating the absence of UV/IR mixing. We also find that the 2-point functions are UV finite with the deformation parameter $\theta$ playing the role of a natural UV cut-off. The possible origin of the absence of UV/IR mixing in noncommutative scalar field theories on $\mathbb{R}^3_\theta$ as well as on $\mathbb{R}^3_\lambda $, another deformation of $\mathbb{R}^3$, is discussed.
\end{abstract}

\newpage
\section{Introduction}
Many of the building blocks of physics fit well with the concepts of Noncommutative Geometry (NCG) \cite{Connes, Connes1}. This may ultimately provide tools to improve our understanding of spacetime at short distance. One argument sometimes put forward is that NCG seems to give a possible way to escape the physical obstruction to the existence of continuous space-time and commuting coordinates at the Planck scale \cite{Doplich1}. This argument has reinforced the interest in Noncommutative Field Theories (NCFT) which started to appear slowly (at least in their modern formulation) from the middle of the 1980's \cite{witt1, mdv1, gm90}. From the beginning of the 2000's, unusual renormalization properties of the NCFT on Moyal spaces $\mathbb{R}^4_\theta$ \cite{minwala} triggered a growing interest, in particular to cope with the UV/IR mixing. This generated many works, leading to the first all order renormalizable scalar field theory with quartic interaction \cite{Grosse:2003aj-pc} where the UV/IR mixing was rendered innocuous through the introduction of a harmonic term. Various properties of this models have been then studied \cite{hghs}-\cite{harald-raimar}. \\

Attempts to accommodate the harmonic term of the above scalar model on $\mathbb{R}^4_\theta$ to a gauge theoretical framework \cite{mdv88-99, cgmw-20} gave rise to a gauge invariant model \cite{Wallet:2007c} bearing formally some properties of a matrix model{\footnote{This model can be viewed as the spectral action stemming from a particular spectral triple \cite{finite-vol} whose noncommutative metric geometries \cite{dand-mar-wal} has been analyzed in \cite{Wallet:2011aa}.}} with a complicated vacuum structure \cite{GWW2}. This precludes the use of standard perturbative treatment in the investigation of renormalizability properties, except on the Moyal plane $\mathbb{R}^2_\theta$ where the gauge invariant model can be related to a 6-vertex model \cite{MVW13}. For reviews, see e.g \cite{dnsw-rev}. Note that the Moyal plane can actually support causal structures stemming from Lorentzian spectral triples, as shown in \cite{causal1}. Renormalizability of the model on $\mathbb{R}^4_\theta$ is still unclear. The same conclusion holds true for alternative approaches based on a IR damping mechanism aiming to neutralize the mixing \cite{Blaschke:2009c}. Note however that the matrix model interpretation of noncommutative gauge theories provided interesting results on (semi-)classical properties and/or one-loop computations. For a review on the related literature, see \cite{matrix2} (see also \cite{matrix3}-\cite{matrix5} and references therein). Recently, scalar field theories on $\mathbb{R}^3_\lambda$, a deformation of $\mathbb{R}^3$ introduced in \cite{Hammaa} (see also \cite{selene}), have been studied in \cite{vit-wal-12}. Some of these NCFT have been shown to be free of perturbative UV/IR mixing \cite{vit-wal-12} and characterized by the occurrence of a natural UV cut-off, stemming from the group algebra structure underlying the $\mathbb{R}^3_\lambda$ algebra \cite{stor-mem}. Gauge theories on $\mathbb{R}^3_\lambda$ have then been investigated \cite{gervitwal-13, GJW-15}. The use of the canonical matrix basis introduced in \cite{vit-wal-12} combined with suitable families of orthogonal polynomials together with a corollary of the spectral theorem \cite{toolkit} permits one to compute the propagator{\footnote{The gauge theories of \cite{gervitwal-13} can be related to (a particular version of) the Alekseev-Recknagel-Schomerus model \cite{ARS} describing a low energy action for brane dynamics on $\mathbb{S}^3$.}}. A gauge-fixing using the suitable machinery of the BRST symmetry \cite{stor-wal} can then be achieved. In \cite{GJW-15}, a family of gauge theories on $\mathbb{R}^3_\lambda$ has been shown to be UV finite to all orders in perturbation and without any IR singularity, among which one particular family of gauge theories was shown to be solvable \cite{stor-mem}.\\

In this paper, we consider the noncommutative space $\mathbb{R}^3_\theta$, a deformation of $\mathbb{R}^3$ for which the star product, hereafter denoted by $\duf$, is closed for the trace functional, namely which satisfies $\tr(f\star_{\mathcal{D}}g)=\tr(f\cdot g)$ for any suitable functions on $\mathbb{R}^3$ $f$ and $g$ (in the notations of Section \ref{section2}). This interesting space has been introduced recently in \cite{KV-15} where it has been shown that the closed star product $\duf$ is linked to the Duflo quantization map \cite{duflo}. The main motivation of \cite{KV-15} was to provide a first attempt to clarify the possible origin of the observed mild perturbative behavior of the NCFT on $\mathbb{R}^3_\lambda$ as stemming either from the particular form of the kinetic operator used in \cite{vit-wal-12}, \cite{gervitwal-13} or being rooted to another yet unidentified property of noncommutative spaces with $\mathfrak{su}(2)$ noncommutativity to which $\mathbb{R}^3_\lambda$ and $\mathbb{R}^3_\theta$ belong. The use of a closed star product enables us to deal with scalar NCFT on $\mathbb{R}^3_\theta$ in which the kinetic operator is the usual Laplacian on $\mathbb{R}^3$. Such NCFT were introduced in \cite{KV-15} and some corresponding classical properties were examined. Here, we study one-loop IR and UV properties of the 2-point function for such real and complex noncommutative scalar field theories. In Section \ref{section2}, we summarize some properties of the Duflo quantization map and the star product $\duf$. The computation of the 2-point functions is given in Section \ref{section3}, supplemented by the appendices \ref{appendixb} and \ref{appendixa}. We find that the 2-point functions for these noncommutative scalar field theories have no IR singularities in the external momenta, indicating the absence of UV/IR mixing. We also find that the 2-point functions are UV finite with the deformation parameter $\theta$ playing the role of a natural UV cut-off. The possible origin of the absence of UV/IR mixing in noncommutative scalar field theories on $\mathbb{R}^3_\theta$ as well as on $\mathbb{R}^3_\lambda $, another deformation of $\mathbb{R}^3$, is discussed in Section {section4}.
\vskip 0,5 true cm
\section{Closed star product and Duflo quantization map.} \label{section2}
For our present purpose, it will be convenient to view the algebra $\mathbb{R}^3_\theta$ as
\begin{equation}
\mathbb{R}^3_\theta:=(\mathcal{M}(\mathbb{R}^3),\duf)\label{algebra},
\end{equation}
where $\mathcal{M}(\mathbb{R}^3)$ is the multiplier algebra of $\mathcal{S}(\mathbb{R}^3)$ (the set of Schwartz functions on $\mathbb{R}^3$) for the star-product $\duf$ defined by
\begin{equation}
f\duf g = \int\frac{d^3k_1}{(2\pi)^3}\frac{d^3k_2}{(2\pi)^3}\widetilde{f}(k_1)\widetilde{g}(k_2)\frac{2|B(k_1,k_2)|\sin(\frac{\theta}{2}|k_1|)\sin(\frac{\theta}{2}|k_2|)}{\theta|k_1||k_2|\sin\left(\frac{\theta}{2}|B(k_1,k_2)|\right)}e^{iB_\mu(k_1,k_2)x^\mu}\label{duflop-full}
\end{equation}
for any $f,g\in\mathcal{S}(\mathbb{R}^3)$ in which the symbol $\widetilde{f}$ denotes generically the Fourier transform of $f$, namely $f(x)=\int\frac{d^3k}{(2\pi)^3}\widetilde{f}(k)e^{ikx}$. In \eqref{duflop-full}, $B_\mu(k_1,k_2)$, $\mu=1,2,3$, stems from the use of the Baker-Hausdorff-Campbell formula occurring naturally in the construction of $\duf$ due to 
the underlying $\mathfrak{su}(2)$ noncommutativity (see below) and, as such, is expressed as an infinite expansion linked to this Lie algebra structure. We will analyze closely this quantity later on. \\

It is instructive to recall the main steps leading to \eqref{duflop-full}, which will fix some notations. The key is to determine a local transformation $T$ acting on the functions of the algebra $\mathbb{R}^3_\theta$, called "gauge transformation" in \cite{KV-15}, with 
\begin{equation}
T=\bbone+\mathcal{O}(\theta)\label{defT},
\end{equation}
where we assume that the deformation parameter $\theta\in\mathbb{R}^+$, such that
\begin{equation}
T(f\duf g)=Tf\star_W Tg\label{equivalenceT},
\end{equation}
where $\star_W$ is the usual Weyl star-product, and determined in such a way that $\duf$ is closed for the trace functional, namely $\tr(f\star_{\mathcal{D}}g)=\tr(f\cdot g)$ or
\begin{equation}
\int d\mu(x)\cdot(f\duf g)(x)=\int d\mu(x)\cdot (f\cdot g)(x)\label{fermeture},
\end{equation}
for any (suitably behaving) functions $f$ and $g$ where ``$\cdot$'' is the commutative product on $\mathbb{R}^3$ and $d\mu(x)$ is some integration measure that can be chosen in the following to be the usual Lebesgue measure (while $\int d\mu(x)\cdot(f\star_W g)(x)\ne\int d\mu(x)\cdot(f\cdot g)(x)$). \\
Such a transformation $T$ may not exist in general. However, its existence is guaranteed by a theorem \cite{FS} for Poisson manifolds with divergenceless bivectors $\omega^{\mu\nu}$, which is the case relevant here. Indeed, the dual of a finite dimensional Lie algebra carries a canonical Poisson manifold structure defined by the Lie Bracket. In the following, we will focus on the simple $\mathfrak{su}(2)$ case for which the conditions of the theorem \cite{FS} applies (in particular, the bivector $\omega^{\mu\nu}(x)=\varepsilon^{\mu\nu\rho}x_\rho)$. Note that Eqn.\eqref{equivalenceT} defines obviously an equivalence relation between $\duf$ and $\star_W$. Besides, $T$ defines a Lie algebra morphism
since \eqref{equivalenceT} implies
\begin{equation}
T([f,g]_{\duf})=[Tf,Tg]_{\star_W}\label{liealgmorph},
\end{equation}
for any $f,g\in\mathbb{R}^3_\theta$. Then, provided a noncommutative structure of the $\mathfrak{su}(2)$ type is obtained from the use of the Weyl map (i.e $[x_\mu,x_\nu]_{\star_W}=i\theta\varepsilon_{\mu \nu}^{\hspace{11pt} \rho}x_\rho$), eqn. \eqref{liealgmorph} combined with the $\theta$ expansion of $T$ \eqref{defT} yields
\begin{equation}
[x_\mu,x_\nu]_{\duf}=[x_\mu,x_\nu]_{\star_W}=
i\theta\varepsilon_{\mu \nu}^{\hspace{11pt} \rho}x_\rho\label{su2per}.
\end{equation}
Hence, \eqref{su2per} shows that the $\mathfrak{su}(2)$ noncommutativity is still compatible with the closed product $\duf$. \\

Such a noncommutative structure can be conveniently implemented by making use of the (poly)differential representation \cite{polydiff-alg} of the operator algebra generated by the standard abstract coordinate operators $\hat{X}_\mu$ satisfying $[\hat{X}_\mu,\hat{X}_\nu]=i\varepsilon_{\mu\nu}^{\hspace{11pt}\rho}\hat{X}_\rho$, $\mu,\nu,\rho=1,2,3$. Let $\pi:\hat{X}_\mu\mapsto\pi(\hat{X}_\mu)=\pi(x_\sigma,\partial_\sigma)$ acting on functions of $\mathcal{M}(\mathbb{R}^3)$ denote this representation, still a Lie algebra morphism. One finds \cite{KV-15}
\begin{equation}
\pi(\hat{X}_\mu)=x_\mu+i\frac{\theta}{2} \varepsilon_{\mu \nu}^{\hspace{11pt} \rho} x_\rho \partial^\nu+
(x_\mu\Delta-x_\nu\partial^\nu\partial_\mu)\Delta^{-1}(\frac{\theta}{2}\sqrt{\Delta}\coth(\frac{\theta}{2}\sqrt{\Delta})-1)\label{polyrep},
\end{equation}
where $\Delta$ is the usual Laplacian on $\mathbb{R}^3$. From now on, we set
\begin{equation}
\pi(\hat{X}_\mu)=\hat{x}_\mu,\ \mu=1,2,3.\label{def-tilde}
\end{equation}
From \eqref{polyrep}, one infers $\hat{x}_\mu\bbone=x_\mu$ so that any symmetrized product verifies $(\hat{x}_{\mu_1}...\hat{x}_{\mu_n})_S\bbone={x}_{\mu_1}...{x}_{\mu_n}$. Then, the general expression for the Weyl quantization map 
\begin{equation}
W(f)=\sum_{k=0}^\infty\frac{1}{k!}(\hat{x}_\mu\partial_\mu)^kf(x)\vert_{x=0} 
\end{equation}
yields
\begin{equation}
W(f)=\int \frac{d^3k}{(2\pi)^3}\widetilde{f}(k)e^{ik\hat{x}}\label{weylmap},
\end{equation}
while it can be easily realized that 
\begin{equation}
W^{-1}(W(f))=W(f)\bbone=f(x). 
\end{equation}
Then, it follows from $f\star_Wg=W^{-1}(W(f)W(g))$ that one can write
\begin{equation}
f\star_Wg=\int\frac{d^3k_1}{(2\pi)^3}\frac{d^3k_2}{(2\pi)^3}\widetilde{f}(k_1)\widetilde{g}(k_2)e^{iB_\mu(k_1,k_2)x_\mu}, \label{weylproducc}
\end{equation}
where we used
\begin{equation}
W^{-1}(e^{ik_1\hat{x}}e^{ik_2\hat{x}})=W^{-1}(e^{iB_\mu(k_1,k_2)\hat{x}_\mu})=e^{iB_\mu(k_1,k_2)x_\mu},\label{BHC}
\end{equation}
stemming from the Baker-Hausdorff-Campbell formula for $\mathfrak{su}(2)$ thus leading to the expression for $B_\mu(k_1,k_2)$ as an infinite expansion. Eqn. \eqref{BHC} implies the following obvious properties
\begin{equation}
B(k,-k)=0,\ B(k,0)=B(0,k)=k,\ B(k_1,k_2)=-B(-k_2,-k_1)\label{Bproperties}.
\end{equation}
Now, from \eqref{equivalenceT} and $f\star_Wg=W^{-1}(W(f)W(g))$, one can write
\begin{equation}
f\duf g=Q^{-1}(Q(f)Q(g))\label{konsduf},
\end{equation}
with
\begin{equation}
Q=W\circ T\label{kquantif},
\end{equation}
while the explicit expression of $T$ computed for the $\mathfrak{su}(2)$ case in \cite{KV-15} is given by
\begin{equation}
T=\frac{2\sinh(\frac{\theta}{2}\sqrt{\Delta})}{\theta\sqrt{\Delta}}\label{Texplicit}.
\end{equation}
By using \eqref{Texplicit} and \eqref{kquantif}, one computes
\begin{equation}
Q(e^{ikx})=\frac{2\sin(\frac{\theta}{2}|k|)}{\theta |k|}e^{ik\hat{x}}\label{planewaves},
\end{equation}
which may be viewed physically as the relation defining the natural plane waves on the noncommutative space $\mathbb{R}^3_\theta$. 
This combined with \eqref{konsduf} yields
\begin{equation}
(f\duf g)(x)=\int\frac{d^3k_1}{(2\pi)^3}\frac{d^3k_2}{(2\pi)^3}\widetilde{f}(k_1)\widetilde{g}(k_2)\big(e^{ik_1x}\duf e^{ik_2x}\big)\label{duflop-gen},
\end{equation}
for any function $f,g\in\mathcal{S}(\mathbb{R}^3)$. It turns out that expressing the closed product $\duf$ under the form \eqref{duflop-gen} will simplify the ensuing computations. \\

Upon using the asymptotic expression of \eqref{duflop-gen} given by
\begin{eqnarray}
(f\duf g)(x)&=&f(x)g(x)+i\frac{\theta}{2}\varepsilon_{\mu \nu}^{\hspace{11pt} \rho}x_\rho\partial^\mu f\partial^\nu g-\theta^2\big[
\frac{1}{8}x_\alpha x_\beta\varepsilon_{\mu \nu}^{\hspace{11pt}\beta}\varepsilon_{\rho \sigma}^{\hspace{11pt}\alpha}\partial^\mu \partial^\rho f\partial^\nu\partial^\sigma g\nonumber\\
&-&\frac{1}{12}x_\sigma\varepsilon_{\mu \nu}^{\hspace{11pt} \sigma}\varepsilon_{\rho \lambda}^{\hspace{11pt} \nu}(\partial^\mu\partial^\rho f\partial^\lambda g+\partial^\mu\partial^\rho g\partial^\lambda f)+\frac{1}{12}\partial_\mu f\partial^\mu g \big]+\mathcal{O}(\theta^3),
\end{eqnarray}
valid for any polynomial functions, one obtains useful relations:
\begin{eqnarray}
[x_\mu,x_\nu]_{\duf}&=&i\theta\varepsilon_{\mu\nu}^{\hspace{11pt}\rho}x_\rho,\label{commutator}\\
x_\mu\duf x_\nu&=&x_\mu x_\nu+i\frac{\theta}{2}\varepsilon_{\mu\nu}^{\hspace{11pt}\rho}x_\rho-\frac{\theta^2}{12}\delta_{\mu\nu}, \label{relation2}\\
x_\mu\duf f&=&x_\mu f+i\frac{\theta}{2}\varepsilon_{\mu\nu}^{\hspace{11pt}\rho}x_\rho\partial^\nu f-\frac{\theta^2}{12}
(x_\mu\partial^2f-x_\nu\partial^\nu\partial_\mu f+\partial_\mu f) \label{relations},
\end{eqnarray}
where, as expected above, \eqref{commutator} is consistent with \eqref{su2per}.\\

Before starting the computation of the 2-point correlation function, some comments are in order:\\
i) First, set $E_k(x):=e^{ikx}$. We remark that the property of the quantization map to be *-algebra morphism implies $E_k(x)^\dag=e^{-ikx}=E_k(-x)$. For now on, we set
\begin{eqnarray}
(e^{ik_1x}\duf e^{ik_2x})(x)&=&\mathcal{W}(k_1,k_2)e^{iB_\mu(k_1,k_2)x_\mu}\nonumber\\
\mathcal{W}(k_1,k_2)&=&\frac{2|B(k_1,k_2)|\sin(\frac{\theta}{2}|k_1|)\sin(\frac{\theta}{2}|k_2|)}{\theta|k_1||k_2|\sin{\frac{\theta}{2}|B(k_1,k_2)|}}\label{Wfunction}.
\end{eqnarray}
According to the remark above, we note that 
\begin{equation}
F_{k_1,k_2}(x):=(e^{ik_1x}\duf e^{ik_2x})(x) 
\end{equation}
satisfies 
\begin{equation}
F_{k_1,k_2}(-x)=(e^{-ik_2x}\duf e^{-ik_1x})(x)=F^\dag_{k_1,k_2}(x), 
\end{equation}
so that changing the sign of the $x$ argument in a product of plane waves amounts to consider the hermitean conjugation of this product.\\
ii) It can be shown \cite{KV-15} that the gauge operator \eqref{Texplicit} can be rewritten as 
\begin{equation}
T=\text{det}^{\frac{1}{2}}\big[\frac{\sinh(\frac{1}{2}\epsilon_\mu\partial^\mu)}{\frac{1}{2}\epsilon_\mu\partial^\mu} \big]
\end{equation}
with $(\epsilon_\mu)_{\nu\rho}=-i\varepsilon_{\mu\nu\rho}$ so that the closed product $\duf$ coincides with the Duflo star-product, i.e the map $Q$ \eqref{kquantif} giving rise to $\duf$ is the Duflo quantization map for $\mathfrak{su}(2)$. For a discussion of this point, see \cite{KV-15} where it is also pointed out that $\duf$ is actually the Kontsevich product \cite{formality}.

\section{Two-point correlation functions for scalar field theories.}\label{section3}

In this section, we will consider real and complex scalar field theories with quartic interactions and both massive Laplacian of $\mathbb{R}^3$ as kinetic operators and study the IR and UV behavior of the corresponding 2-point correlation functions at the one-loop order. By a simple inspection of the perturbative expansion of the generating functional for the correlation functions, it can be easily realized that the one-loop 2-point correlation function for the real field case receives two types of contributions, hereafter called Type-I and Type-II contributions, depending whether or not both the contracted lines giving rise to the propagator are related to two consecutive exponential factors (as it is apparent in \eqref{interaction} below), upon taking into account the cyclicity of the trace $\int d^3x$. \\
As it can be expected in the complex scalar field case, the form of the interaction term determines which type of contributions should be taken into account for the 2-point function: only Type-I contributions matter when the interaction is $\int d^3x\ \Phi^\dag\duf\Phi\duf\Phi^\dag\duf\Phi$ (hereafter called invariant interaction) while both Type-I and Type-II actually contribute when the interaction is $\int d^3x\ \Phi^\dag\duf\Phi^\dag\duf\Phi\duf\Phi$ (hereafter called non-invariant interaction), which is an obvious consequence of the form of the interaction. Invariance or non-invariance is with respect to the transformations defined by the natural action of the automorphisms of the algebra viewed as a (right-)module on itself, compatible with the canonical hermitean structure used here, namely $h(a_1,a_2)=a^\dag_1a_2$. Thus, one can write $\Phi^g=g\duf \Phi$, for any $g$ with $g^\dag\duf g=g\duf g^\dag=\bbone$ so that $h(\Phi_1^g,\Phi_2^g)=h(\Phi_1,\Phi_2)$. \\
In Subsection \ref{subsection31}, we will first consider the real field case and first focus on the analysis of Type-I contributions, showing that they are IR and UV finite. The extension to the case of complex scalar field with interaction 
$\int d^3x\ \Phi^\dag\duf\Phi\duf\Phi^\dag\duf\Phi$ leading to the conclusion that the related 2-point function is also UV and IR finite is then given. In Subsection \ref{subsection33}, we go back to the real field case and consider Type-II contributions. These are found to be IR finite. The corresponding UV behavior is then analyzed. The case of complex model with non-invariant interaction is also discussed.

\subsection{Scalar field theories and Type-I contributions.}\label{subsection31}
We first consider a real-valued scalar field theory with quartic interaction whose classical action is
\begin{equation}
S=\int d^3x\big[\frac{1}{2}\partial_\mu\phi\duf\partial_\mu\phi+\frac{1}{2}m^2\phi\duf\phi+\frac{\lambda}{4!}\phi\duf\phi\duf\phi\duf\phi\big]\label{real-clas-action},
\end{equation}
where $\duf$ is the closed product defined above. In \eqref{real-clas-action}, the fields and parameters are assumed to have the usual $\mathbb{R}^3$ mass dimensions, namely 
\begin{equation}
[\phi]=\frac{1}{2},\ [\lambda]=1,\ [m]=1. \label{massdim}
\end{equation}
UV (resp. IR) regime is defined, as usual, by the region of large (resp.) small momenta. Notice that in the present case, the kinetic term of \eqref{real-clas-action} simplifies as $\int d^3x\ (\partial_\mu\phi\partial_\mu\phi+m^2\phi\phi)$ thanks to the fact that the product $\duf$ is closed \cite{KV-15}. This permits one to avoid the (often) difficult step of the computation of the propagator in noncommutative field theories which here is nothing but the usual massive propagator of a scalar field. Besides, the formal commutative limit of the kinetic term is obvious.\\
The interaction term can be conveniently recast into the form
\begin{eqnarray}
S_{int}&=&\frac{\lambda}{4!}\int d^3x\int\ \left[\prod_{i=1}^4 \frac{d^3k_i}{(2\pi)^3}
\widetilde\phi(k_i)\right](e^{ik_1x}\duf e^{ik_2x}\duf e^{ik_3x}\duf e^{ik_4x})(x)\label{interaction}\\
&=&\frac{\lambda}{4!}\int\ \left[\prod_{i=1}^4 \frac{d^3k_i}{(2\pi)^3}
\widetilde\phi(k_i)\right]\mathcal{W}(k_1,k_2)\mathcal{W}(k_3,k_4)\delta(B(k_1,k_2)+B(k_3,k_4))\label{interaction1}.
\end{eqnarray}
We will use alternatively \eqref{interaction} and \eqref{interaction1} in the computation of the contributions to the 2-point correlation function. Notice that the standard conservation law of the momenta $\delta(\sum_{i=1}^4k_i)$ of the commutative $\phi^4$ theory on $\mathbb{R}^3$ in replaced by a non-linear one as it can be seen from the delta function in \eqref{interaction1}. This complicates strongly the perturbative calculations, which however can be partly overcome by a suitable use of \eqref{interaction} combined with properties of the plane waves and cyclicity of the trace.\\

A typical contribution of Type-I to the one-loop effective action is easily found to be given by
\begin{equation}
\Gamma^{(I)}_2=\int d^3x \left[\prod_{i=1}^4 \frac{d^3k_i}{(2\pi)^3}\right]\widetilde\phi(k_3)\widetilde\phi(k_4)\frac{\delta(k_1+k_2)}{k_1^2+m^2}(e^{ik_1x}\duf e^{ik_2x}\duf e^{ik_3x}\duf e^{ik_4x})(x)\label{type1}
\end{equation}
where we dropped the overall constant $\sim\lambda$. Combining \eqref{type1} with \eqref{Bproperties}, \eqref{Wfunction} and
\begin{equation}
(e^{ikx}\duf e^{-ikx})(x)=\frac{4}{\theta^2}\frac{\sin^2(\frac{\theta}{2}|k|)}{|k|^2}\label{plane-norm}
\end{equation}
we obtain
\begin{eqnarray}
\Gamma^{(I)}_2&=&\int d^3x\frac{d^3k_3}{(2\pi)^3}\frac{d^3k_4}{(2\pi)^3}\widetilde\phi(k_3)\widetilde\phi(k_4)(e^{ik_3x}\duf e^{ik_4x})(x)
\omega^{(I)}\nonumber\\
&=&\int d^3x(\phi\duf\phi)(x)\omega^{(I)}=\int d^3x\ \phi(x)\phi(x)\omega^{(I)}\label{amplitude-I}
\end{eqnarray}
with
\begin{equation}
\omega^{(I)}=\frac{4}{\theta^2}\int\frac{d^3k}{(2\pi)^3}\frac{\sin^2(\frac{\theta}{2}|k|)}{k^2(k^2+m^2)} \label{type1omega}.
\end{equation}
The integral over the internal momentum $k$ is finite since
\begin{equation}
 \omega^{(I)}=\frac{4}{\theta^2}\int\frac{d^3k}{(2\pi)^3}\frac{\sin^2(\frac{\theta}{2}|k|)}{k^2(k^2+m^2)}=\frac{1}{\pi^2\theta^{2}}\int^{\infty}_{0}dr\frac{1-\cos(\theta r)}{r^2+m^2}=\frac{1-e^{-\theta m}}{2m\pi\theta^2}\label{type1-final}.\end{equation}
A similar result holds true obviously for the other Type-I contributions. \\

Hence, Type-I contributions are UV finite and do not exhibit IR singularity. We conclude that whenever $\theta\ne0$, Type-I contributions cannot generate UV/IR mixing. Notice that the closed star product structure of the quadratic part of the effective action still survives the one-loop quantum corrections at it is apparent from \eqref{amplitude-I}.

From \eqref{type1-final}, one readily obtain the small $\theta$ expansion of $\omega^{(I)}$, namely
\begin{equation}
\omega^{(I)}_{\theta\rightarrow 0}=\Lambda+...\label{comutlim1},
\end{equation}
where the ellipsis denote finite ($\mathcal{O}(1)$) contributions and $\Lambda=\frac{1}{2\pi\theta}$. Thus, we recover as leading divergent term the expected linear divergence (showing up when $\Lambda\to\infty$) which occurs in the 2-point function for the commutative theory with $\Lambda=\frac{1}{2\pi\theta}$ as the  UV cutoff. In physical words, the present noncommutativity of $\mathfrak{su}(2)$ type gives rise to a natural UV cutoff for the scalar field theory \eqref{real-clas-action}. Notice that this holds true even when $m=0$ as it can be seen from \eqref{type1-final}. Note also that $[\Lambda]=1$ since $[\theta]=-1$.\\

{\bf{Complex scalar field theories.}}\par

The above one-loop analysis extends easily to Type-I contributions for the 2-point function of the complex scalar field theories with invariant or non-invariant interactions. \\
In the case of invariant interaction, the 2-point function only receives Type-I contributions. The action is
\begin{equation}
S=\int d^3x\big[\partial_\mu\Phi^\dag\duf\partial_\mu\Phi+m^2\Phi^\dag\duf\Phi+
{\lambda}\Phi^\dag\duf\Phi\duf\Phi^\dag\duf\Phi\big]\label{complx-clas-action}.
\end{equation}
By using a standard perturbative expansion, one easily find that a typical contribution to the one-loop quadratic part of the effective action is given by
\begin{equation}
({\Gamma_{\mathbb{C}}}^{(I)})_2=\int d^3x \left[\prod_{i=1}^4 \frac{d^3k_i}{(2\pi)^3}\right]\widetilde{\Phi^\dag}(k_3)\widetilde{\Phi}(k_4)\frac{\delta(k_1+k_2)}{k_1^2+m^2}(e^{ik_1x}\duf e^{ik_2x}\duf e^{ik_3x}\duf e^{ik_4x})(x)\label{type1-complx}
\end{equation}
where we dropped again the overall constant $\sim\lambda$. From the analysis of Subsection \ref{subsection31}, one obtains immediately
\begin{equation}
({\Gamma_{\mathbb{C}}}^{(I)})_2=\int d^3x\ \Phi^\dag(x)\Phi(x)\ \omega_{\mathbb{C}}^{(I)},\label{omega-complx}
\end{equation}
with 
\begin{equation}
\omega_\mathbb{C}^{(I)}\sim\omega^{(I)}
\end{equation}
and $\omega^{(I)}$ given by \eqref{type1omega}. $\omega_\mathbb{C}^{(I)}\sim\omega^{(I)}$ is again finite since relations similar to \eqref{type1-final}-\eqref{comutlim1} still holds true for $\omega_\mathbb{C}^{(I)}$. \\
A similar conclusion obviously holds true for the Type-I contributions involved in the 2-point function related to the scalar theory with non-invariant interaction. However, Type-II contributions mentioned at the beginning of this section are also involved. These will be examined in the next subsection.\par

\subsection{Type-II contributions.}\label{subsection33}

Let us go back to the real scalar field theory \eqref{real-clas-action}. A typical Type-II contribution to the one-loop effective action is given by
\begin{equation}
\Gamma^{(I \hspace{-1pt} I)}_2=\int\frac{d^3k_2}{(2\pi)^3}\frac{d^3k_4}{(2\pi)^3}\widetilde\phi(k_2)\widetilde\phi(k_4)\omega^{(I \hspace{-1pt} I)}(k_2,k_4)\label{type2}
\end{equation}
with
\begin{eqnarray}
\omega^{(I \hspace{-1pt} I)}(k_2,k_4)&=&\int d^3x\frac{d^3k_1}{(2\pi)^3}\frac{d^3k_3}{(2\pi)^3}\frac{\delta(k_1+k_3)}{k_1^2+m^2}(e^{ik_1x}\duf e^{ik_2x}\duf e^{ik_3x}\duf e^{ik_4x})(x)\nonumber\\
&=&\int d^3x\frac{d^3k}{(2\pi)^3}\frac{1}{k^2+m^2}(e^{ikx}\duf e^{ik_2x}\duf e^{-ikx}\duf e^{ik_4x})(x)\label{type2-omega}
\end{eqnarray}
where now the internal momentum involves two non neighboring exponential factors.\\

We first consider the Infrared regime of \eqref{type2-omega} corresponding to the small external momenta region, i.e $k_2\sim0,\ k_4\sim0$. From \eqref{Wfunction}, one infers
\begin{equation}
(e^{ik_1x}\duf e^{ik_2x})_{|k_2=0}(x)=e^{ik_1x}\label{expo-unit}
\end{equation}
which simply reflects the fact that $e^{ik\hat{x}}_{|k=0}=\bbone$. Then, one can write
\begin{eqnarray}
\omega^{(I \hspace{-1pt} I)}(0,k_4)&=&\int d^3x\frac{d^3k}{(2\pi)^3}\frac{1}{k^2+m^2}(e^{ikx}\duf e^{-ikx}\duf e^{ik_4x})(x)\nonumber\\
&=&\delta(k_4)\frac{4}{\theta^2}\int\frac{d^3k}{(2\pi)^3}\frac{\sin^2(\frac{\theta}{2}|k|)}{k^2(k^2+m^2)}=\delta(k_4)\ \omega^{(I)}\label{ir-omega2}.
\end{eqnarray}
where $\omega^{(I)}$ is given by \eqref{type1-final} and we used \eqref{plane-norm} to obtain the second equality. From the discussion for the Type-I contributions given in Subsection \ref{subsection31}, we conclude that \eqref{ir-omega2} is not IR singular (and also UV finite). A similar result holds true for $\omega^{(I \hspace{-1pt} I)}(k_2,0)$. The extension to complex scalar field theories is obvious.\\

From this and the discussion of Subsection \ref{subsection31}, we conclude that no IR singularity shows up in the 2-point functions for the real and complex (even massless $m=0$) scalar field theories at one-loop so that these NCFT are free from UV/IR mixing.\\

Unfortunately, since exponentials no longer simplify in \eqref{type2-omega}, now one has to deal with infinite expansions stemming from the $B(k_1,k_2)$ function in \eqref{Wfunction} and/or delta functions with non-linear arguments which complicate considerably the UV analysis of the Type-II contributions. However, this situation can be slightly simplified by considering a somewhat restricted situation for which the coordinate functions $x_\mu$ satisfying the relation \eqref{su2per} for the Lie algebra $\mathfrak{su}(2)$ are represented as Pauli matrices. Note that a similar representation is used in models related to quantum gravity \cite{q-grav} in which are involved noncommutative structures similar to the one considered here. Namely, introduce the following morphism of algebra $\rho: \mathfrak{su}(2)\to \mathbb{M}_2(\mathbb{C})$
\begin{equation}
\rho(\hat{x}_\mu) = \theta \sigma_\mu,\ \rho(\mathbb{I}) = \mathbb{I}_2\label{decadix}.
\end{equation}
where $\hat{x}_\mu$, $\mu=1,2,3$ is defined in \eqref{def-tilde}. From the usual properties of the Pauli matrices, one obtains the following relation
\begin{equation} 
\rho (\hat{x}_i \hat{x}_j) = \rho(\hat{x}_i) \rho(\hat{x}_j) = \theta^2 \delta_{ij} \mathbb{I}_2 + i \frac{\theta}{2} \varepsilon_{ij}^{\hspace{5pt} k} \rho (\hat{x}_k)\label{representation_operateur}
\end{equation}
which will give rise to a rather simple expression for the exponential factors occurring in \eqref{type2-omega}. Notice that \eqref{representation_operateur} is consistent with (\ref{relation2}) as it can be seen by using $Q=W \circ T$ on $x_\mu\star_\mathcal{D} x_\nu$. By using $T = 1 + \frac{\theta^2}{24} \Delta + \frac{\theta^4}{16*5!} \Delta^2 + \mathcal{O}(\theta^6)$, we easily obtain
\begin{eqnarray}
T(x_\mu \star_\mathcal{D} x_\nu) &=& T(x_\mu x_\nu) + i \frac{\theta}{2} \varepsilon_{\mu\nu}^{\hspace{5pt} \rho} T(x_\rho) - \frac{\theta^2}{12} \delta_{\mu\nu} T(1)\nonumber\\
&=& (x_\mu x_\nu + \frac{\theta^2}{12}\delta_{\mu\nu}) + i \frac{\theta}{2} \varepsilon_{\mu\nu}^{\hspace{5pt} \rho} x_\rho - \frac{\theta^2}{12} \delta_{\mu\nu}.
\end{eqnarray}
Then we can write 
\begin{equation}
Q(x_\mu \star_\mathcal{D} x_\nu) = \hat{x}_\mu \hat{x}_\nu = W(x_\mu x_\nu) + i \frac{\theta}{2} \varepsilon_{\mu\nu}^{\hspace{5pt} \rho} W(x_\rho),
\end{equation}
so that 
\begin{equation}
\hat{x}_\mu \hat{x}_\nu = \frac{1}{2}(\hat{x}_\mu \hat{x}_\nu + \hat{x}_\nu \hat{x}_\mu) + i \frac{\theta}{2} \varepsilon_{\mu\nu}^{\hspace{5pt} \rho} \hat{x}_\rho. 
\end{equation}
Applying $\rho$ \eqref{decadix} and $\sigma_\mu\sigma_\nu+\sigma_\nu\sigma_\mu = 2\delta_{\mu\nu}$ yields the result.\\

From \eqref{representation_operateur} we obtain after some algebraic manipulations
\begin{equation}
 e^{ik^\mu \rho(\hat{x}_\mu)} = \cos\left(\theta |k| \right) \mathbb{I}_2 + i \frac{\sin\left(\theta |k| \right)}{\theta |k|} k^\mu \rho(\hat{x}_\mu)\label{expo-coupe}.
\end{equation}
This implies
\begin{equation}
[ e^{ik^\mu\rho(\hat{x}_\mu)} , e^{ip^\nu \rho(\hat{x}_\nu)}] = - i \theta \frac{\sin\left(\theta |k| \right)}{\theta |k|} \frac{\sin\left(\theta |p| \right)}{\theta |p|} \varepsilon_{\sigma \nu}^{\hspace{8pt} \rho} k^\sigma p^\nu \rho(\hat{x}_\rho)\label{commut-expon}.
\end{equation}

Now $\Gamma_2^{(I \hspace{-1pt} I)}$ \eqref{type2} can be conveniently rewritten as 
\begin{equation} 
\Gamma^{(I \hspace{-1pt} I)}_2 = \Gamma^{(I)}_2 + \int \frac{d^3k_2 }{(2\pi)^3} \frac{d^3k_4 }{(2\pi)^3} \widetilde{\phi}(k_2) \widetilde{\phi}(k_4) I(k_2,k_4)\label{TypeII_asI}
\end{equation}
where in obvious notations
\begin{eqnarray}
 I(k_2,k_4) &=& \int \frac{d^3k }{(2\pi)^3} \frac{d^3x}{k^2+m^2} \left[ e^{ik^\nu x_\nu} , e^{ik_2^\nu x_\nu} \right]_{\duf} \duf e^{-ik^\nu x_\nu} \duf e^{ik_4^\nu x_\nu} \nonumber \\
 &=& \left(\frac{2}{\theta}\right)^4 \int \frac{d^3k }{(2\pi)^3} \frac{d^3x}{k^2+m^2} \left(\frac{\sin(\frac{\theta}{2}|k|)}{|k|}\right)^2 \frac{\sin(\frac{\theta}{2}|k_2|)}{|k_2|} \frac{\sin(\frac{\theta}{2}|k_4|)}{|k_4|}\nonumber\\
 &\times& \mathcal{Q}^{-1} \left( \left[ e^{ik^\mu \hat{x}_\mu} , e^{ik_2^\nu \hat{x}_\nu} \right] e^{-ik^\sigma \hat{x}_\sigma} e^{ik_4^\rho \hat{x}_\rho} \right) \label{ik2k4}.
\end{eqnarray}
By further making use of \eqref{decadix} together 
with \eqref{commut-expon}, we arrive after a lengthy computation given in Appendix \ref{appendixb} to the following expression
\begin{equation} 
I(k_2,k_4) = \frac{J(k_2,k_4)}{\pi^3 \theta^4} \int d\alpha d\beta d r \frac{\sin^2(\frac{\theta}{2}r)}{r^2+m^2} \left[ \frac{1}{2} \sin\left(2 \theta r \right) \sin\gamma + \sin^2\left(\theta r \right) \sin^2\frac{\gamma}{2} \right] \sin\alpha \label{typeII_spherique},
\end{equation}
with
\begin{equation}
 J(k_2,k_4) = \frac{\sin\left(\theta |k_2| \right) \sin(\frac{\theta}{2}|k_2|)}{|k_2|} u^\mu \delta_\mu^{'}(k_4)\label{france-j},
\end{equation}
in which we used spherical coordinates for the momentum $k$, namely $k=(r=|k|,\alpha,\beta)$, $u_\mu$ is the $\mu$-component of a unit vector $u$, $\gamma$ is the angle between the momenta $k$ and $k_2$ (depending only on $\alpha$, $\beta$ and $\alpha_2$, $\beta_2$ entering the spherical coordinates for $k_2$) and $\delta^\prime_\mu$ is defined by $\langle \delta^\prime_\mu, f\rangle=-\frac{\partial f}{\partial k_{4}^\mu}$ for any test function $f$. One can check that \eqref{typeII_spherique} is finite, as shown in Appendix \ref{appendixb} and already conclude that Type-II contributions are UV finite.\\
The radial integration in \eqref{typeII_spherique} can be performed by further using 
\begin{equation}\begin{split}
\int_0^\infty \frac{\cos(ax)}{\beta^2+x^2} dx &=\frac{\pi}{2\beta}e^{-a\beta}, \, a \geq 0, \, \text{Re}(\beta)>0, \\ 
\int_0^\infty \frac{\sin(ax)}{\beta^2+x^2} dx &= \frac{1}{2\beta} \left[ e^{-a \beta} \overline{\text{Ei} (a \beta)} - e^{a \beta} \text{Ei} (- a \beta) \right] \, , a>0 \, , \beta >0\label{GS-integral-prime}
\end{split}\end{equation} 
where $\text{Ei}$ is the exponential integral function defined by
\begin{equation}
 \text{Ei}(x) = - \underset{e \rightarrow 0^{+}}{\lim} \left[ \int^{-e}_{-x} \frac{e^{-t}}{t}dt + \int^\infty_e \frac{e^{-t}}{t}dt\right],\ x>0
\end{equation}
and one has
\begin{equation}
 \text{Ei}(x) = \textbf{C} + \ln |x| +\sum \limits^\infty_{n=1} \frac{x^n}{n.n!},\ x\neq 0,
\end{equation}
in which \textbf{C} the Euler-Mascheroni constant. We obtain
\begin{eqnarray}
 \frac{1}{2}  \int d r \frac{\sin^2(\frac{\theta}{2}r)}{r^2+m^2} \sin\left(2 \theta r \right) &=& \frac{1}{16m} \left[ 2e^{-2 \theta m} \overline{\text{Ei} (2 \theta m)} - 2e^{2 \theta m} \text{Ei} (-2 \theta m) - e^{-3 \theta m} \overline{\text{Ei} (3 \theta m)} \right. \nonumber \\
 & & + \left. e^{3 \theta m} \text{Ei} (-3 \theta m) - e^{- \theta m} \overline{\text{Ei} ( \theta m)} + e^{ \theta m} \text{Ei} (- \theta m) \right] \label{commutative_typeII_1},
\end{eqnarray}
and
\begin{equation} 
\int d r \frac{\sin^2(\frac{\theta}{2}r)}{r^2+m^2} \sin^2\left(\theta r \right) = \frac{\pi}{8m} \left[ 1-\left(1+\sinh(\theta m)\right)e^{-2\theta m} \right]\label{commutative_typeII_2}.
\end{equation}
In the small $\theta$ limit (i.e formal commutative limit $\theta\to0$), one infers
\begin{equation}
\frac{J(k_2,k_4)}{\pi^3 \theta^4} = \frac{1}{\pi^3 \theta^2} |k_2| u^n \delta^{'}_n(k_4) + \mathcal{O}(\theta^0), \label{decadix-1}
\end{equation}
\begin{equation}
 \text{(\ref{commutative_typeII_2})} =\frac{\pi}{8} \theta + \mathcal{O}(\theta^2)
\end{equation}
and
\begin{equation}
\text{(\ref{commutative_typeII_1})}=(6\ln3-8\ln2)\theta+\mathcal{O}(\theta^2)\label{decadix-2}.
\end{equation}
Combining \eqref{decadix-1}-\eqref{decadix-2} with \eqref{typeII_spherique} yields the following small $\theta$ limit:
\begin{equation}
I(k_2,k_4)=\frac{C(\alpha_2,\beta_2)}{\theta} |k_2| u^n \delta^{'}_n(k_4) + \mathcal{O}(\theta^0),
\end{equation}
where $C(\alpha_2,\beta_2)$ is finite. Hence, as for the Type-I contributions, the $\theta$ expansion of $I(k_2,k_4)$ is 
\begin{equation}
I\sim \Lambda + ...
\end{equation}
where the ellipsis still denote finite $\mathcal{O}(1)$ contributions and $\Lambda=\frac{1}{2\pi\theta}$. Thus, combining this result with the decomposition (\ref{TypeII_asI}) of $\Gamma^{(I \hspace{-1pt} I)}_2$, we recover one more time the expected linear divergence when $\Lambda\to\infty$ ($\theta\to0$) occurring in the 2-point function for the commutative theory. Again, the present $\mathfrak{su}(2)$ noncommutativity generates a natural UV cutoff for the scalar field theory. Notice that this holds true even when $m=0$. This result extends obviously to complex scalar field theories.

\section{Discussion and Conclusion.}

We have considered the noncommutative space $\mathbb{R}^3_\theta$ defined in \eqref{algebra} for which the star product $\duf$ is closed for the trace functional $\int d^3x$. We have investigated one-loop properties of real or complex, massive or massless, scalar field theories with quartic interactions on $\mathbb{R}^3_\theta$ for which the kinetic part is the usual Laplacian on $\mathbb{R}^3$ which can be obtained thanks to the closeness property of $\duf$. Hence, the formal commutative limit of these NCFT yields usual $\phi^4$-type theories on $\mathbb{R}^3$. One-loop contributions to the 2-point functions are computed. We found that the 2-point functions for these NCFT do not involve IR singularities in the external momenta, even in the massless case. This signals the absence of UV/IR mixing of the type occurring in NCFT on Moyal spaces. The UV behavior of the 2-point functions is also considered. We found that the 2-point functions are UV finite with the deformation parameter $\theta$ providing a natural UV cut-off $\Lambda=\frac{1}{2\pi\theta}$. In the commutative $\theta\to0$ limit, we recover the usual UV linear divergence of the 2-point function.\\

We note that the absence of UV/IR mixing has also been shown for various NCFT on $\mathbb{R}^3_\lambda$, another space still having $\mathfrak{su}(2)$-type noncommutativity with however a different star product. These NCFT are scalar fields theories on $\mathbb{R}^3_\lambda$ with quartic interaction \cite{vit-wal-12} as well as gauge theory models on $\mathbb{R}^3_\lambda$ characterized by different classical vacua, studied respectively in \cite{gervitwal-13} and \cite{GJW-15}, for which each related kinetic operator is different from the present Laplacian of $\mathbb{R}^3$. In each case, the absence of UV/IR mixing can be viewed as resulting mainly from the existence of a natural (UV) cut-off linked to the very structure of $\mathbb{R}^3_\lambda$ together with the choice of a ``reasonably decaying" propagator in the far UV region, hence not rooted in a specific form for the kinetic operator{\footnote{This is particularly apparent in \cite{GJW-15} where quantum stability of the vacuum for the family of gauge theories, which does not hold in \cite{gervitwal-13}, is supplemented by perturbative finiteness to all orders, one among these gauge theories exhibiting a relationship to integrable 2-d Toda hierarchies \cite{stor-mem}. While UV finiteness renders an actual discussion of the mixing superfluous, the way the cut-off shows up in the amplitudes of diagrams is very clear in these latter theories.}}. From this, we conclude that the absence of UV/IR mixing in NCFT with $\mathfrak{su}(2)$-type noncommutativity reflects mainly the structure of $\mathbb{R}^3_\theta$ and $\mathbb{R}^3_\lambda$ as deformation of $\mathbb{R}^3$ with star product giving rise to the defining relation for an $\mathfrak{su}(2)$ Lie algebra between coordinates. This, by the way, provides an answer to the question, raised in \cite{KV-15},   of the origin of the mild perturbative behavior of the NCFT on deformed $\mathbb{R}^3$.\\

The origin of a natural cut-off in both $\mathbb{R}^3_\theta$ and $\mathbb{R}^3_\lambda$ spaces seems very likely rooted in the existence of a common structure to these algebras related to the convolution algebra of the group $SU(2)$, denoted hereafter by $\mathcal{A}^0(SU(2))=(L^2(SU(2)),\bullet)$. Here, 
\begin{equation}
(f\bullet g)(u)=\int_{SU(2)} d\mu(t)f(ut^{-1}) g(t),
\end{equation}
 for any functions $f,g\in L^1(SU(2))$ where $d\mu(t)$ is the $SU(2)$ Haar measure. Now, introduce $\widehat{\mathbb{R}^3_\lambda}=\bigoplus_{j\in\frac{\mathbb{N}}{2}}(\mathbb{M}_{2j+1}(\mathbb{C}),\cdot)$, where the symbol ``$\cdot$'' is the matrix (operator) product. This is the operator representation of $\mathbb{R}^3_\lambda=(\mathcal{M}(\mathbb{R}^3),\star_\lambda)$ given in \cite{vit-wal-12} together with $\star_\lambda$ from which the elements of the adapted basis for $\mathbb{R}^3_\lambda$ are obtained as symbols of the canonical matrix basis for $\widehat{\mathbb{R}^3_\lambda}$ through a suitable representation map, says, $\rho:{\mathbb{R}^3_\lambda}\to\widehat{\mathbb{R}^3_\lambda}$. It turns out that $\widehat{\mathbb{R}^3_\lambda}$ can be identified with (the $SU(2)$ Fourier transform of) $\mathcal{A}^0(SU(2))$ with isomorphism defined by $\mathcal{F}:\mathcal{A}^0(SU(2))\to\widehat{\mathbb{R}^3_\lambda}$ with 
\begin{equation}
\check{f}:=\mathcal{F}(f)=\bigoplus_{j\in\frac{\mathbb{N}}{2}}\int_{SU(2)}d\mu(x)f(x)t^j(x^{-1}),
\end{equation}
 for any $f\in L^2(SU(2))$ together with the inverse map $\mathcal{F}^{-1}:\widehat{\mathbb{R}^3_\lambda}\to \mathcal{A}^0(SU(2))$ with 
\begin{equation}
\mathcal{F}^{-1}(\check{f})(x)=\bigoplus_{j\in\frac{\mathbb{\mathbb{N}}}{2}}(2j+1)\mbox{tr}_j(t^j(x)\check{f}).
\end{equation}
 Here, $\mathcal{F}$ is the $SU(2)$ Fourier transform, and $t^j(x)$ is the matrix of the coefficients of the irreducible representation $\chi_j$ for $x\in SU(2)$ defined (in the notations of e.g \cite{vit-wal-12}) by $(t^j(x))_{mn}=\langle jm|\chi_j(x)|jn\rangle$ which is a Wigner $D$-matrix and $\mbox{tr}_j$ is the canonical trace on $\mathbb{M}_{2j+1}(\mathbb{C})$. Hence, the decomposition of $\widehat{\mathbb{R}^3_\lambda}$ as an orthogonal sum which reflects the Peter-Weyl decomposition of $L^2(SU(2))$ transfers to $\mathbb{R}^3_\lambda$. This generates the appearance of a cut-off in the amplitudes of the diagrams in \cite{vit-wal-12, gervitwal-13, GJW-15}, thanks to a general factorization property of the partition function. Now, from the construction of the star product $\duf$ recalled in Section \ref{section2}, one has ${Q}^{-1}:\widehat{\mathbb{R}^3_\lambda}\to \mathbb{R}^3_\theta$. Hence, the Peter-Weyl decomposition of $\widehat{\mathbb{R}^3_\lambda}$ should transfer to $\mathbb{R}^3_\theta$ defined in \eqref{algebra} so that the corresponding partition function should also obey a factorization property as the one above leading the occurrence of a cut-off in the amplitudes for the diagrams. Note that a suitable basis for $\mathbb{R}^3_\theta$ should presumably be obtained from the image by $Q^{-1}$ of the canonical basis of $\widehat{\mathbb{R}^3_\lambda}$. It would be interesting to characterize all the mathematical properties of the maps relating these various deformations of $\mathbb{R}^3$ and the convolution algebra $\mathcal{A}^0(SU(2))$ or its Fourier transform $\widehat{\mathbb{R}^3_\lambda}$, this latter being at the crossroad of various models pertaining to Brane physics (see \cite{ARS}, \cite{gervitwal-13}), 2+1 quantum gravity (see \cite{q-grav} and ref. therein) and NCFT. We will come to this point in a forthcoming publication.

\vskip 1 true cm
{\bf{Acknowledgements:}} We thank P. Vitale for correspondence on ref. \cite{KV-15} at early stage of this work. JCW warmly thanks N. Franco and F. Latr\'emoli\`ere for various discussions on group $C^*$-algebras and related topics in connection with the present work. TJ thanks S. Meljanac, Z \v{S}koda and A. Samsarov for several discussions in the early stages of this work. The authors would like to acknowledge the support of the  RBI-T-WINNING project funded by European Commission under H2020. The work by TJ has been partially supported by Croatian Science Foundation under the project (IP-2014-09-9582).

\appendix

\section{Appendix A.}\label{appendixb}
We start from
\begin{eqnarray}
 I(k_2,k_4) &=&
 \left(\frac{2}{\theta}\right)^4 \int \frac{d^3k }{(2\pi)^3} \frac{d^3x}{k^2+m^2} \left(\frac{\sin(\frac{\theta}{2}|k|)}{|k|}\right)^2 \frac{\sin(\frac{\theta}{2}|k_2|)}{|k_2|} \frac{\sin(\frac{\theta}{2}|k_4|)}{|k_4|}\nonumber\\
& &\times\mathcal{Q}^{-1} \left( \left[ e^{ik^\mu \hat{x}_\mu} , e^{ik_2^\nu \hat{x}_\nu} \right] e^{-ik^\sigma \hat{x}_\sigma} e^{ik_4^\rho \hat{x}_\rho} \right) \label{ik2k4-1}.
\end{eqnarray}
Then we use \eqref{decadix}-\eqref{expo-coupe} to write
\begin{equation} 
 \left[ e^{ik\hat{x}} , e^{ik_2 \hat{x}} \right] e^{-ik \hat{x}} e^{ik_4 \hat{x}} \overset{\rho}{\longmapsto} - i \theta \frac{\sin\left(\theta |k| \right)}{\theta |k|} \frac{\sin\left(\theta |k_2| \right)}{\theta |k_2|} \varepsilon_{\mu \nu}^{\hspace{8pt} \rho} k^\mu k_2^\nu \rho(\hat{x}_\rho) e^{-ik \rho(\hat{x})} e^{ik_4 \rho(\hat{x})} \label{representationtype2}
\end{equation}
where :
\begin{eqnarray}
 \rho(\hat{x}_\rho) e^{-ik \rho(\hat{x})} &=& \cos\left(\theta |k| \right) \rho(\hat{x}_\rho) - i \frac{\sin\left(\theta |k| \right)}{\theta |k|} k^\sigma \rho(\hat{x}_\rho)\rho(\hat{x}_\sigma) \nonumber \\
 &=& -\theta^2 \frac{\sin\left(\theta |k| \right)}{\theta |k|} k_\rho \mathbb{I}_2 + \left[ \cos\left(\theta |k| \right) \delta^\nu_\rho + \frac{\theta}{2} \frac{\sin\left(\theta |k| \right)}{\theta |k|} \varepsilon_{\rho \mu}^{\hspace{8pt} \nu} k^\mu \right] \rho(\hat{x}_\nu) \label{interm-2}.
\end{eqnarray}
After some algebraic manipulations, the RHS of \eqref{representationtype2} can be cast into the simpler form
\begin{equation}
- i\theta \frac{\sin\left(\theta |k| \right)}{\theta |k|} \frac{\sin\left(\theta |k_2| \right)}{\theta |k_2|} \left[\cos\left(\theta |k| \right) \varepsilon_{\mu \nu}^{\hspace{11pt} \rho} k^\mu k_2^\nu + \frac{\sin\left(\theta |k| \right)}{2|k|} \left(|k|^2 k_2^\rho - k_\sigma k_2^\sigma  k^\rho \right) \right] \rho(\hat{x}_\rho) e^{ik_4 \rho(\hat{x})}\label{interm-1}.
\end{equation}
Then combining \eqref{representationtype2}-\eqref{interm-1} with the action of \eqref{decadix} on \eqref{ik2k4-1} yields
\begin{equation} 
 I (k_2,k_4) = - \frac{i}{\theta} \left(\frac{2}{\theta}\right)^4 \int \frac{d^3k }{(2\pi)^3} \frac{d^3x}{k^2+m^2} A^\rho(k,k_2) \mathcal{Q}^{-1} \left[ \rho(\hat{x}_\rho) e^{ik_4 \rho(\hat{x})} \right] \frac{\sin(\frac{\theta}{2}|k_4|)}{|k_4|}\label{finale-1}
\end{equation}
where
\begin{eqnarray}
A^\rho(k,k_2) &=& \frac{\sin^2(\frac{\theta}{2}|k|)}{|k|^2} \bigg[\cos(\theta |k| ) \frac{\sin(\theta |k| )}{|k|} \varepsilon_{\mu \nu}^{\hspace{11pt} \rho} k^\mu k_2^\nu\nonumber\\
& &+ \frac{\sin^2(\theta |k| )}{2|k|^2} (|k|^2 k_2^\rho - k_\sigma k_2^\sigma  k^\rho )\bigg] \frac{\sin(\theta |k_2| ) \sin(\frac{\theta}{2}|k_2|)}{|k_2|^2}\label{finale-1bis}.
\end{eqnarray}
Then we write \eqref{ik2k4-1} as 
\begin{equation}
I (k_2,k_4) = \frac{1}{\theta } \left(\frac{2}{\theta}\right)^3 \int \frac{d^3k }{(2\pi)^3} \frac{d^3x}{k^2+m^2} A^\rho(k,k_2) \left(\frac{\partial}{\partial k_4^\rho} e^{ik_4 x} \right).\label{qualityf}
\end{equation}
Integrating over $x$, the last term between parenthesis in \eqref{qualityf} gives the derivative of the Dirac distribution $\delta_\mu^{'}(k_4)$ defined, for any test function $\psi$, by $<\delta_\mu^{'} , \psi> = - \left. \frac{\partial \psi}{\partial k_4^\mu} \right\vert_{k_4=0}$.\\
Finally, we introduce $u^\mu$ the unit vector in the direction $\mu$ and $\gamma$ the angle between the two momenta $k$ and $k_2$ and we have
\begin{eqnarray}
 A^\rho(k,k_2) &=& \left. \frac{\sin^2(\frac{\theta}{2}|k|)}{|k|^2} \right[\cos(\theta |k| ) \sin(\theta |k| ) \sin\gamma \nonumber\\
 & &+ \left. \sin^2(\theta |k| ) \frac{1-\cos\gamma}{2} \right] \frac{\sin(\theta |k_2| ) \sin(\frac{\theta}{2}|k_2|)}{|k_2|} u^\rho\label{arho},
\end{eqnarray}
which combined with \eqref{qualityf} produces
\begin{equation} 
I (k_2,k_4) = \frac{J(k_2,k_4)}{\pi^3 \theta^4} \int d\alpha d\beta d r \frac{\sin^2(\frac{\theta}{2}r)}{r^2+m^2} \left[ \frac{1}{2} \sin\left(2 \theta r \right) \sin\gamma + \sin^2\left(\theta r \right) \sin^2\frac{\gamma}{2} \right] \sin\alpha \label{typeII_spherique-bis}
\end{equation}
with
\begin{equation}
 J(k_2,k_4) = \frac{\sin\left(\theta |k_2| \right) \sin(\frac{\theta}{2}|k_2|)}{|k_2|} u^\rho \delta_\rho^{'}(k_4)\label{france-j-bis}
\end{equation}
which coincide with \eqref{typeII_spherique} and \eqref{france-j} respectively.
In (\ref{typeII_spherique-bis}), we have decomposed $d^3k$ into the spherical coordinates $k = (r=|k|,\alpha,\beta)$ for which the angle $\gamma$ depends only in the angles $\alpha$, $\beta$ that define $k$ and the angles $\alpha_2$, $\beta_2$ that define $k_2$. Thus, the two integrations over $\alpha$ and $\beta$ are finite. \\

One can easily show that the integration over $k$ is finite. Indeed, one has
\begin{equation}
\int d\alpha d\beta dr \frac{\sin^2(\frac{\theta}{2}r)}{r^2+m^2} \left[ \frac{1}{2} \sin\left(2 \theta r \right) \sin\gamma + \sin^2\left(\theta r \right) \sin^2\frac{\gamma}{2} \right] \sin\alpha \leqslant \int_0^\infty \frac{dr}{r^2+m^2} < \infty.
\end{equation}
Hence, we conclude that the Type-II contribution \eqref{typeII_spherique-bis} is UV and IR finite (even for $m=0$ see Subsection \ref{subsection33}).

\section{Appendix B.}\label{appendixa}
In this appendix, we present a computation of Type-I contributions using the framework introduced in \cite{KV-15} for which the closed form for the quantity $B(k_1,k_2)$ relies on the use of \eqref{decadix}. In this case, $B(k_1,k_2)$ takes the closed form given by
\begin{equation}
B(k_1,k_2)=\frac{2\arcsin|p_1 \oplus p_2|}{\theta}\big(\frac{p_1 \oplus p_2}{|p_1 \oplus p_2|}\big),\ p_i=\frac{k_i}{|k_i|}{\sin{(\frac{\theta}{2}|k_i|)}},\ i=1,2,\label{B-cut}
\end{equation}
stemming from the combination of \eqref{BHC}, \eqref{decadix} and \eqref{expo-coupe}, where we defined 
\begin{equation}
p \oplus q = \sqrt{1-\left\vert q \right\vert^2} p + \sqrt{1-\left\vert p \right\vert^2} p - p \times q. 
\end{equation}
Having a closed expression for $B(k_1,k_2)$ allows for a direct computation by exploiting the delta functions appearing in the expressions for the 2-point functions which however is complicated by the appearence of Jacobian stemming from the fact that the argument of the delta functions are non linear which, physically, may be interpreted as non-linear momenta conservation law. Of course, the resulting expression agrees with the one obtained in Subsection \ref{subsection31}.\\
From \eqref{interaction1}, a typical Type-I contribution can be written as
\begin{eqnarray}
\Gamma^{(I)} & = & (2\pi)^3 \int \bigg[\prod \limits_{a=1}^4 \frac{d^3k_a}{(2\pi)^3}\bigg] \widetilde{\phi}(k_1) \widetilde{\phi}(k_2) \frac{\mathcal{W}(k_1,k_2) \mathcal{W}(k_3,k_4)}{k_3^2+m^2} \delta(k_3+k_4) \delta \left(B(k_1,k_2)+B(k_3,k_4)\right) \nonumber \\
& = & \int  \frac{d^3k_1}{(2\pi)^3} \frac{d^3k_2}{(2\pi)^3} \frac{d^3k}{(2\pi)^3} \widetilde{\phi}(k_1) \widetilde{\phi}(k_2) \frac{\mathcal{W}(k_1,k_2) \mathcal{W}(k,-k)}{k^2+m^2} \delta \left(B(k_1,k_2)+B(k,-k)\right) \label{calctypeI}
\end{eqnarray}
By using $B(k,-k)=0$ and $\mathcal{W}(k,-k) = \left(\frac{2 \sin(\frac{\theta}{2} |k|)}{\theta |k|} \right)^2$, a standard computation yields
\begin{equation} 
\Gamma^{(I)} = \int\frac{d^3k_1}{(2\pi)^3} \frac{d^3k_2}{(2\pi)^3} \widetilde{\phi}(k_1) \widetilde{\phi}(k_2)  \omega^{(I)} (k_1,k_2),\label{gamma-pferd}
\end{equation}
with
\begin{equation} \label{IntTypeI}
\omega^{(I)} (k_1,k_2) = \left(\frac{2}{\theta}\right)^2 \left( \int \frac{d^3k}{(2\pi)^3} \frac{\sin^2 \left(\frac{\theta}{2}|k|\right)}{k^2(k^2+m^2)} \right) \frac{W(k_1,k_2)}{\left\vert \frac{\partial B}{\partial v} (u,v) \right\vert}_{\substack{u=k_1 \\ v=k_2}} \delta\left(k_1+k_2\right),
\end{equation}
in which $\left\vert \frac{\partial B}{\partial v}\right\vert$ is a Jacobian. To obtain \eqref{gamma-pferd}, \eqref{IntTypeI}, use has been made of 
\begin{equation*}
\int d^3 x f(x) \delta(g(x)) = \int d^3 y f(g^{-1}(y)) \left\vert \frac{\partial g}{\partial x} \right\vert^{-1} \delta(y) = f\left(g^{-1}(0)\right) \left\vert \frac{\partial g}{\partial x} \right\vert^{-1}_{x=g^{-1}(0)}.
\end{equation*}
We have already shown that $\int \frac{d^3k}{(2\pi)^3} \frac{\sin^2 \left(\frac{\theta}{2}|k|\right)}{k^2(k^2+m^2)}$ is finite in Subsection \ref{subsection31}. \\

To compute the Jacobian, we write it as the product of the two determinants $\left\vert \frac{\partial q}{\partial v} \right\vert$ and $\left\vert\frac{\partial B}{\partial q}\right\vert$, with $q = \frac{v}{|v|} \sin\left(\frac{\theta}{2} |v| \right)$. The first determinant can be expressed as: 
\begin{equation} \label{jacob1}
\left\vert \frac{\partial q}{\partial v} \right\vert = \frac{\theta}{4 v^2} \sin\left(\frac{\theta}{2}|v| \right)\sin(\theta |v|).
\end{equation} 
For the other one, we compute explicitely the components of the matrix $\frac{\partial B_\mu}{\partial q_\nu}$ for arbitrary momenta and then compute the determinant in the special case $u=-v$.\\ 
From \eqref{B-cut}, we get :
\begin{eqnarray}
\frac{\partial B_\mu}{\partial q_\nu}(p,q) & = & \frac{2}{\theta \left\vert p \oplus q \right\vert} \left[ \arcsin(\left\vert p \oplus q \right\vert) \left( \sqrt{1-|p|^2}\delta_{\nu\mu}-\frac{q_\nu p_\mu}{\sqrt{1-|q|^2}} - \varepsilon_{\sigma\nu\mu}p^\sigma \right) + \right.\nonumber \\
& + & \left( \frac{1}{\sqrt{1-|p \oplus q|}}-\frac{\arcsin(|p \oplus q|)}{|p \oplus q|} \right) \left( (1-|p|^2)q_\nu + \sqrt{(1-|p|^2)(1-|q|^2)}p_\nu - \right. \nonumber \\
& & \hspace{3.5cm} - \left. \left. (p.q) \left(p_\nu + \sqrt{\frac{1-|p|^2}{1-|q|^2}} q_\nu \right) \right) \frac{(p \oplus q)_\mu}{\left\vert p \oplus q \right\vert} \right],
\end{eqnarray}
which leads, for $u=-v$ that is for $p=-q$, to :
\begin{equation} \label{jacob2}
\left\vert \frac{\partial B}{\partial q}\right\vert_{q=-p} = \left( \frac{2}{\theta} \right)^3 \frac{1}{\sqrt{1-|p|^2}}.
\end{equation}
Finally, combining (\ref{jacob1}) and (\ref{jacob2}) with (\ref{IntTypeI}), for $k_1 = - k_2 = s$, we obtain :
\begin{equation} \label{typeI}
\omega^{(I)} (s,-s) = \left(\frac{2}{\theta}\right)^2 \int \frac{d^3k}{(2\pi)^3} \frac{\sin^2 \left(\frac{\theta}{2}|k|\right)}{k^2(k^2+m^2)},
\end{equation}
which is finite for non-zero $\theta$ and agrees with the result of Subsection \ref{subsection31}.


\end{document}